\documentclass[twocolumn]{openjournal}
\usepackage{graphicx} % Required for inserting images

\usepackage[colorlinks,linkcolor=blue,citecolor=blue,urlcolor=blue ]{hyperref}

\usepackage[T1]{fontenc}

\begin{document}
\title{Breaking up with the continuous exoplanet mass-radius relation\vspace{-3em}}
% The list of authors, and the short list which is used in the headers.
% If you need two or more lines of authors, add an extra line using \newauthor
\author{Kathryn Edmondson}
\author{Jordan Norris}
\author{Eamonn Kerins$^{*}$}
\affiliation{Department of Physics and Astronomy, University of Manchester, Oxford Road, Manchester M13 9PL, UK.}
\thanks{$^*$Corresponding author: Eamonn.Kerins@manchester.ac.uk}

% Abstract of the paper
\begin{abstract}
We use a carefully selected sub-sample of 1053 confirmed exoplanets from the NASA Exoplanet Archive to construct empirical power-law exoplanet mass-radius-temperature ($M$-$R$-$T$) relations. Using orthogonal distance regression to account for errors in both mass and radius, we allow the data to decide: 1) the number of distinct planetary regimes; 2) whether the boundaries of these regimes are best described by broken power laws joined at mass break points, or by discontinuous power laws motivated by changes in equations of state and temperature. We find strong support from the data for three distinct planetary $M$-$R$ regimes and for those regimes to be discontinuous. Our most successful model involves an $M$-$R$-$T$ relation in which ice/rock (rocky) and ice-giant (neptunian) planets are segregated by a pure-ice equation of state, whilst neptunes and gas giant (jovian) planets are segregated by a mass break at $M_{\rm br} = 115\pm19~M_{\oplus}$. The rocky planet regime is shown to follow $M \propto R^{0.34\pm0.01}$, whilst neptunes have $M\propto R^{0.55\pm0.02}$. Planets in both regimes are seen to extend to similar maximum masses. In the jovian regime we find that $M \propto R^{0.00\pm0.01}T^{0.35\pm 0.02}$, where $T$ is the planet equilibrium temperature. This implies that, for jovian planets detected so far, equilibrium temperature alone provides a robust estimator of mass. 
\end{abstract}

\maketitle

\section{Introduction}

Understanding the relationship between exoplanet mass and radius is crucial to constraining internal composition and testing planet formation simulations, as well as predicting the detectability of a planet to aid in the design of future surveys. Some studies \citep[e.g.][]{Seager2007,Swift2012} take the physically-motivated approach of modelling planetary composition using equations of state to infer a mass-radius relation. Others take an empirical approach, involving applying analytic, probabilistic, or machine-learning models to data \citep[e.g.][]{Otegi2020,ChenKipping2017,Mousavi-Sadr2023}. \cite{Weiss2013} defined two planetary regimes either side of $150 ~M_{\oplus}$, and fit a mass-radius-incident-flux ($M$-$R$-$F$) plane to each population. The small and large planet regimes were found to follow $R \propto M^{0.53}F^{-0.03}$ and $R \propto M^{0.039}F^{0.094}$ respectively. \cite{Wolfgang2016} focused on small planets less than $4 R_{\oplus}$ and used a Hierarchical Bayesian model to obtain a probabilistic model with $M \propto R^{1.3}$. A later relation developed by \cite{Bashi2017} again split exoplanets into two regimes, this time with a floating break-point. This yielded $R \propto M^{0.55}$ and $R \propto M^{0.01}$ for the small and large planet regimes, and a break-point at $127 M_{\oplus}$, which was attributed to the mass at which electrons in hydrogen become degenerate.

\cite{ChenKipping2017} used a similar methodology to \cite{Wolfgang2016}, extending their analysis to develop a probablistic model, \texttt{Forecaster}, that also included brown dwarfs, small stars, and solar system dwarf planets and moons. They identified four regimes: terran worlds, for which $R \propto M^{0.28}$; neptunian worlds, for which $R \propto M^{0.59}$; jovian worlds, for which $R \propto M^{-0.04}$ and stellar worlds, for which $R \propto M^{0.88}$. The transition between terran and neptunian worlds occurred at $2 M_{\oplus}$, and between neptunian and jovian worlds at $130 M_{\oplus}$.

Power laws are often used for exoplanet mass-radius ($M$-$R$) relations. However, \citet{Ning2018} argued that such a model is not adequately flexible to describe more complex features in a $M$-$R$ diagram and, hence, they presented a non-parametric model through the use of Bernstein polynomials.

\cite{Ulmer-Moll2019} investigated the dependence of radius on other parameters, such as equilibrium temperature, semi-major axis and properties of the host star through a machine-learning approach. This method circumvents the need to classify planets, or use differing $M$-$R$ relations in different regimes, and was better able to characterise the spread of hot-jupiter radii than \cite{ChenKipping2017}.

All of the models discussed so far imposed the condition that a $M$-$R$ relation should be continuous, however \cite{Otegi2020} took a different approach by categorising exoplanets below $120 M_{\oplus}$ as rocky or volatile-rich according to an equation of state of pure water. Their results were unaffected by the exact equation of state and temperature assumption used, yielding $R \propto M^{0.29}$ for rocky planets and $R \propto M^{0.63}$ for those rich in volatiles. The main downside of a discontinuous $M$-$R$ relation is that it permits relations that overlap in mass or radius, resulting in non-uniqueness. But, this may allow a more accurate characterisation of underlying discontinuities that arise as a result of planet compositional transitions.

The large scatter in the radii of massive planets cannot be explained simply by a deterministic power law and is generally attributed to atmospheric bloating due to stellar insolation. \cite{Enoch2012} investigated which parameters contribute to this scatter for Saturn-, Jupiter- and high-mass planets and found that the radius of a Jupiter-mass planet could be predicted from equilibrium temperature and semi-major axis, with no mass dependence at all. Saturn- and high-mass planet radii were found to depend also on stellar metallicity and planet mass, however the division of exoplanets into these three populations is somewhat arbitrary. \cite{Thorngren2018} and \cite{Sestovic2018} both used hierarchical Bayesian modelling to investigate the cause of bloating and concluded that insolation is key in characterising inflated radii, favouring a $M$-$R$-$F$ relation. \cite{Thorngren2018} also found that radius inflation decreased for high enough temperatures, deducing that the mechanism by which bloating occurs is Ohmic dissipation, in which a planet's magnetic field interacts with atmospheric flows, thereby transferring heat to the planet interior. Ohmic dissipation initially has a greater impact with increasing equilibrium temperature and thus ionisation, however for very high temperatures the atmospheric winds are subject to magnetic drag and the process becomes inhibited \citep{Batygin2011}.

There have been several studies to investigate whether the composition of rocky planets can be described by the abundances of rock-building elements in the host star. \cite{Schulze2021} found that in general, the mass and radius of a rocky exoplanet is consistent with a model of the planet derived from the $\frac{\mathrm{Fe}}{\mathrm{Mg}}$ and $\frac{\mathrm{Si}}{\mathrm{Mg}}$ ratios found in the host star's photosphere. However, there are mechanisms after planet formation that can alter planet composition. For example, Mercury has been vastly enriched in iron, potentially due to mantle-stripping collisions.

In this paper we seek to take advantage of the rapid expansion of exoplanet data to revisit the exoplanet $M$-$R$ relation. By using a carefully selected data subsample, we allow the data alone to decide on the required number and location of the different planetary regimes. We also test for the support between continuous and discontinuous $M$-$R$ relations, including extension to a $M$-$R$-$T$ relation for massive planets. The paper is organized as follows. In Section~\ref{data} we define our data subsample and in section~\ref{piecewise} we use the data to construct piece-wise power laws, allowing for floating planet mass break points and accounting for errors in both mass and radius. In section~\ref{modifications}, we explore massive-planet $M$-$R$-$T$ relations as well as discontinuous $M$-$R$ relations for lower-mass planets. In Section~\ref{comparison} we compare the performance of our fits to each other and to the widely-used probabilistic \texttt{Forecaster} code \citep{ChenKipping2017}. In Section~\ref{conclusions} we present our conclusions.

\section{Data} \label{data}

Exoplanet data were retrieved from the NASA Exoplanet Archive\footnote{\url{https://exoplanetarchive.ipac.caltech.edu/}, accessed during September 2022}, consisting at the time of 5171 confirmed exoplanets. Of these, 1057 have both mass and radius measurements quoted with uncertainties. In cases where a planet had multiple measurements of the same parameter, we chose the entry that optimized the fractional uncertainty in density. In cases where equilibrium temperature measurements were available, the most precise measurement was taken. While previous works have implemented significance cuts that require the mass and radius to exceed a given multiple of their errors, we do not consider such a selection cut here due to its potential for bias in the small planet regime \citep{Burt2018}.

We impose two plausibility criteria: firstly, a planet must have a mass less than the minimum mass for the onset of deuterium burning; secondly, a planet must have a density less than that of pure iron.
In the first criterion, there is some ambiguity as to the exact mass above which deuterium burning occurs. \cite{Spiegel2011} have demonstrated that this mass is not well defined in general, due to the influence of several factors including helium abundance, initial deuterium abundance and metallicity; leading to deuterium burning limits between around $11 M_J$ and $16 M_J$, depending on model assumptions. For our sample selection we adopt the canonical value of $13 M_J$. 

Some planets in the sample give a bulk density larger than that of pure iron. While measurement error may be a cause, it has been suggested by \cite{Mocquet2014} that there could be a physical explanation for these anomalous cases. \cite{Mocquet2014} demonstrated that the bare core of a gas giant which has been stripped of its atmosphere during migration could have been irreversibly compressed, leading to a new regime of very high density exoplanets. However, there has been no follow-up work to date and we consider the current sample of very high density exoplanets to be too small to warrant their treatment as a new sub-population of planets. Consequently, super-iron density planets are excluded by our plausibility criteria.

The radius of a pure-iron planet, $R_{\rm Fe}$, has been obtained from \cite{Fortney2007} by setting the rock mass fraction for a rock/iron planet to be zero, yielding
\begin{eqnarray}
R_{\rm Fe} = 0.0975 (\log M)^2 + 0.4938 \log M + 0.7932,
\label{iron_radius}
\end{eqnarray}
where both mass and radius are in Earth units. An equivalent expression may be found for the mass of a pure iron planet, $M_{\rm Fe}$, by rearranging Equation~\ref{iron_radius} and recasting $M \rightarrow M_{\rm Fe}$ and $R_{\rm Fe} \rightarrow R$ to give
\begin{eqnarray}
\log M_{\rm Fe} = -2.532 + 5.128 \sqrt{0.3900 R - 0.0655},
\label{iron_mass}
\end{eqnarray}
where, again, quantities are in Earth units.

To quantify our plausibility criteria, we define a weighting factor for a measurement to be the probability that an exoplanet satisfies our plausibility criteria, assuming that the true parameter is distributed as a Gaussian such that $R_{\rm t} \sim N(R, \sigma _R^2)$, where $R_{\rm t}$ is the true radius, $R$ is the measured radius with uncertainty $\sigma _R$, and $N$ is the normal distribution. Similarly, the true mass $M_{\rm t}$ is assumed to follow $M_{\rm t} \sim N(M, \sigma _M^2)$, where $M$ is the measured mass with uncertainty $\sigma_M$. The weighting factor for radius, $W_{R}$, is therefore
\begin{eqnarray}
W_{R} = P(R_{\rm t} \geq R_{\rm Fe}),
\label{radius_weight}
\end{eqnarray}
where $R_{\rm Fe}$ is obtained by substituting $M$ into Equation~\ref{iron_radius}. To account for both a high-density and high-mass planet, the total mass weighting factor, $W_{M}$, is given by
\begin{eqnarray}
W_{M} = P(M_{\rm t} \leq M_{\rm Fe}) \cdot P(M_{\rm t} \leq 13 M_J),
\label{mass_weight}
\end{eqnarray}
where $M_{\rm Fe}$ is obtained by substituting $R$ into Equation~\ref{iron_mass}. To embed this information in a single variable, we define a combined weighting factor to be the product of $W_{R}$ and $W_{M}$. Four planets received a combined weighting factor of effectively zero and so were excluded from the sample, leaving a total of 1053 exoplanets to be considered in this analysis. The weighting factors are carried forward and considered in the mass-radius relations developed in this paper.

\section{Piece-wise mass-radius relations} \label{piecewise}

One of the simplest deterministic models for a $M$-$R$ relation is a power law, which can be written as 
\begin{eqnarray}
\label{power_law}
\frac{R}{R_{\oplus}} = k  \left(\frac{M}{M_{\oplus}} \right) ^\beta,
\end{eqnarray}
where $k$ is a constant and $\beta$ is the index of the power law. In this section we consider piece-wise power-law models. If we impose that each piece $n$ joins to form a continuous relation, then such models reduce to a simple $y=m_nx+c$ form in log space, where
\begin{eqnarray}
\label{piecewise_function}
y = m_1 x + c + \sum_{i=2}^{n} (m_{i-1} - m_i)b_{i-1},
\end{eqnarray}
where $m_n$ is the gradient of the $n^{th}$ piece, $c$ is the intercept of the first piece and $b_n$ is the $n^{th}$ break-point, provided that $b_{n-1} \leq x < b_n$ for $n \geq 1$ with $b_0 = 0$. We treat $m_n$, $b_n$ and $c$ as free parameters of the model, and we use orthogonal distance regression (ODR) to fit $y = \log R$ as a function of $x = \log M$ to Equation~\ref{piecewise_function}, accounting for errors in both quantities. We also incorporate the weighting factors derived in section~\ref{data} into our analysis by combining them with the statistical weights from the measurement errors, such that
\begin{eqnarray}
\label{weight}
W_{{\rm tot}, X} = W_X \cdot \frac{1}{\sigma _{X}^2},
\end{eqnarray}
where $X$ represents mass or radius, and $W_X$ is the result from Equations~\ref{radius_weight} or \ref{mass_weight}. $W_{{\rm tot}, X}$ in Equation~\ref{weight} is the weight used in the ODR fitting routine, hence data points may have small weights stemming from large errors, or a small probability of satisfying our plausibility criteria, or both. These data points therefore carry low importance to the fit.

\begin{figure}
\includegraphics[width=0.45\textwidth]{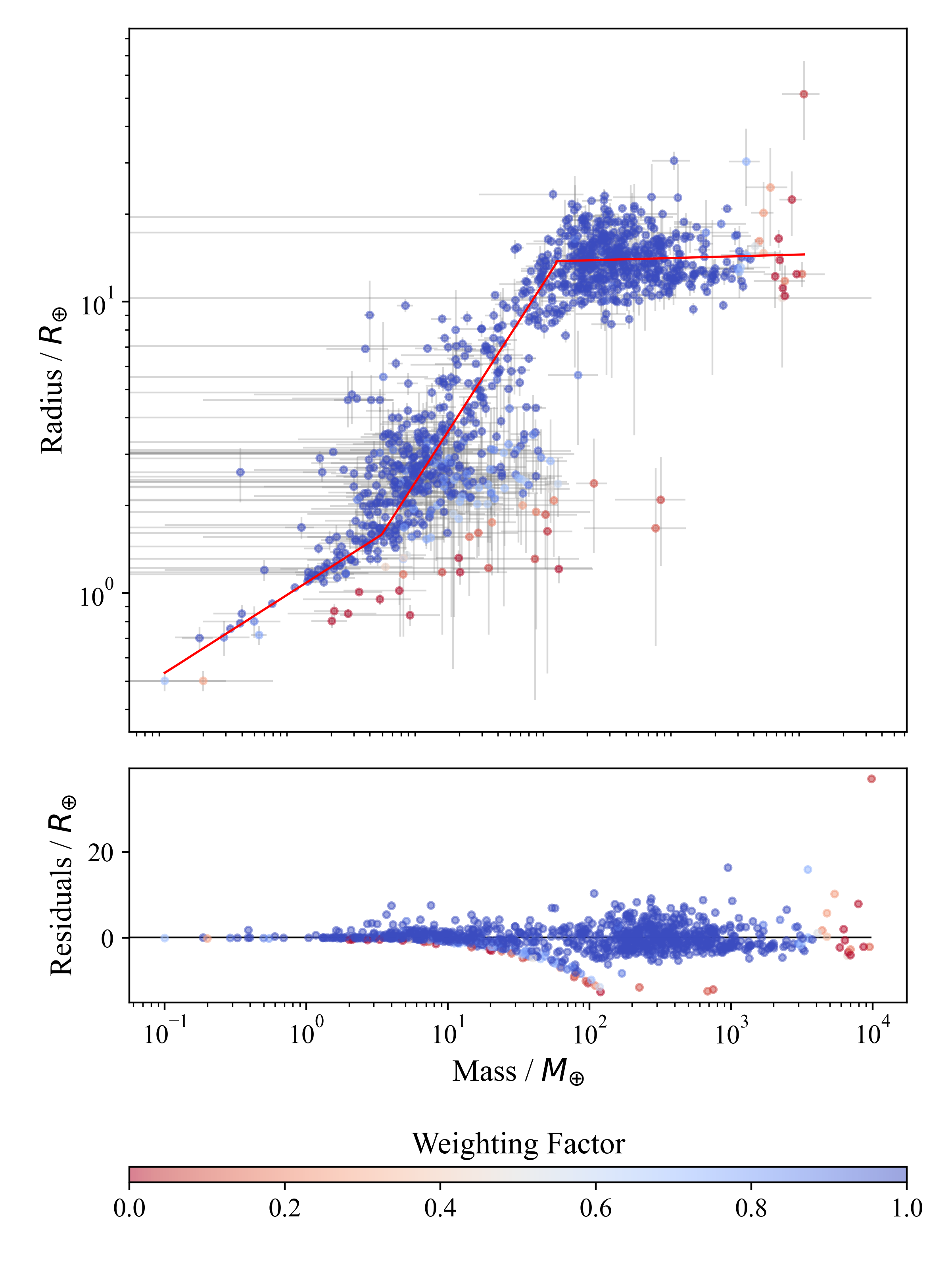}
\caption{A log-log plot of exoplanet radius against mass, in Earth units. The colours represent the combined weighting factor such that blue planets satisfy our plausibility criteria completely, whereas red planets do not. The red line shows the three-piece function fitted by ODR, and the residuals from the model are plotted in the lower panel. The average error-normalised absolute difference between model and measured radius is 7.64. Note the significant residual tail that curves downward at around $50~M_{\oplus}$.}
\label{3_piece} 
\end{figure}

From visual inspection of a $M$-$R$ diagram (e.g. Figure~\ref{3_piece}), it is clear that a single power law is not optimal to describe the data. The cases in which $n=2$ and $n=3$ both yield reasonable fits, however for the case of $n=4$, the last break-point was found to be at a mass greater than the largest mass in the data set. We interpret this result as an indication that no more than three distinct regimes are supported by current data.

We use planet radius to measure the success of a model, by defining the metric
\begin{equation}
{\cal M} \equiv \left\langle \frac{|R_{o} - R_{e}|}{\sigma}\right\rangle, 
\label{r_metric}
\end{equation}
where $R_o$ is the observed radius, $R_e$ is the radius expected by the model given the measured mass, and $\sigma$ is the radius measurement uncertainty. The choice to use the prediction of radius, rather than mass, as the basis for the metric is a pragmatic one, stemming from a clear insensitivity of radius to mass for the most massive planets that is driven by underlying physics rather than by measurement uncertainty. We find that a two-piece model gives a value of ${\cal M} = 7.81$, whereas a three-piece model gives ${\cal M} = 7.64$. For a continuous piece-wise $M$-$R$ relation current data prefer, though not strongly prefer, three rather than two distinct exoplanetary regimes, which we label here onwards as rocky, neptunian and jovian. The best-fit three-piece continuous model is presented in Figure~\ref{3_piece}.

The transitions between regimes are found to be at $4.95 \pm 0.81 M_{\oplus}$ and $115 \pm 19 M_{\oplus}$, and the power law parameters from Equation~\ref{power_law} are as follows: $k = 1.01 \pm 0.03$ and $\beta = 0.28 \pm 0.03$ in the rocky regime; $k = 0.53 \pm 0.05$ and $\beta = 0.68 \pm 0.02$ in the neptunian regime; and $k = 13.0 \pm 1.2$ and $\beta = 0.012 \pm 0.003$ in the jovian regime. From the residuals, planets in the rocky regime appear to be well constrained by a power law, however in the neptunian regime there is a noticeable systematic downturn indicative of a population of planets that coherently deviates from the rest of the neptunian regime. This is evidence that a continuous $M$-$R$ may not be appropriate, and that the rocky and neptunian planetary regimes overlap \citep[e.g.][]{Otegi2020}. We consider a discontinuous model in Section~\ref{disc_mrt_model}.

\section{Beyond a continuous mass-radius relation} \label{modifications}

\subsection{Temperature Considerations} \label{temp_inclusion}

Many hot jupiters show evidence of radius bloating due to high levels of stellar insolation. This motivates a need to include temperature by considering mass-radius-temperature $M$-$R$-$T$ relations for the jovian regime. In order to include a temperature dependence, we extend the use of ODR to fit a plane in mass-radius-temperature log space, accounting for errors in all measured quantities. We carry forward the same analysis of weighting factors that was performed in section~\ref{piecewise}, resulting in a multiplicative power law of the form
\begin{eqnarray}
\label{product_power}
\frac{R}{R_{\oplus}} = C T_{\rm eq}^{\beta_1} \left( \frac{M}{M_{\oplus}} \right)^{\beta_2},
\label{eq:mrt}
\end{eqnarray}
where $C$, $\beta_1$ and $\beta_2$ are constants and $T_{\rm eq}$ is the equilibrium temperature of the planet. For the rocky and neptunian regimes, and for jovian planets without an equilibrium temperature measurement, a simple $M$-$R$ power law is used to predict their radii, relaxing the constraint that the global $M$-$R$ relation must be continuous.

\begin{figure}
\includegraphics[width=0.45\textwidth]{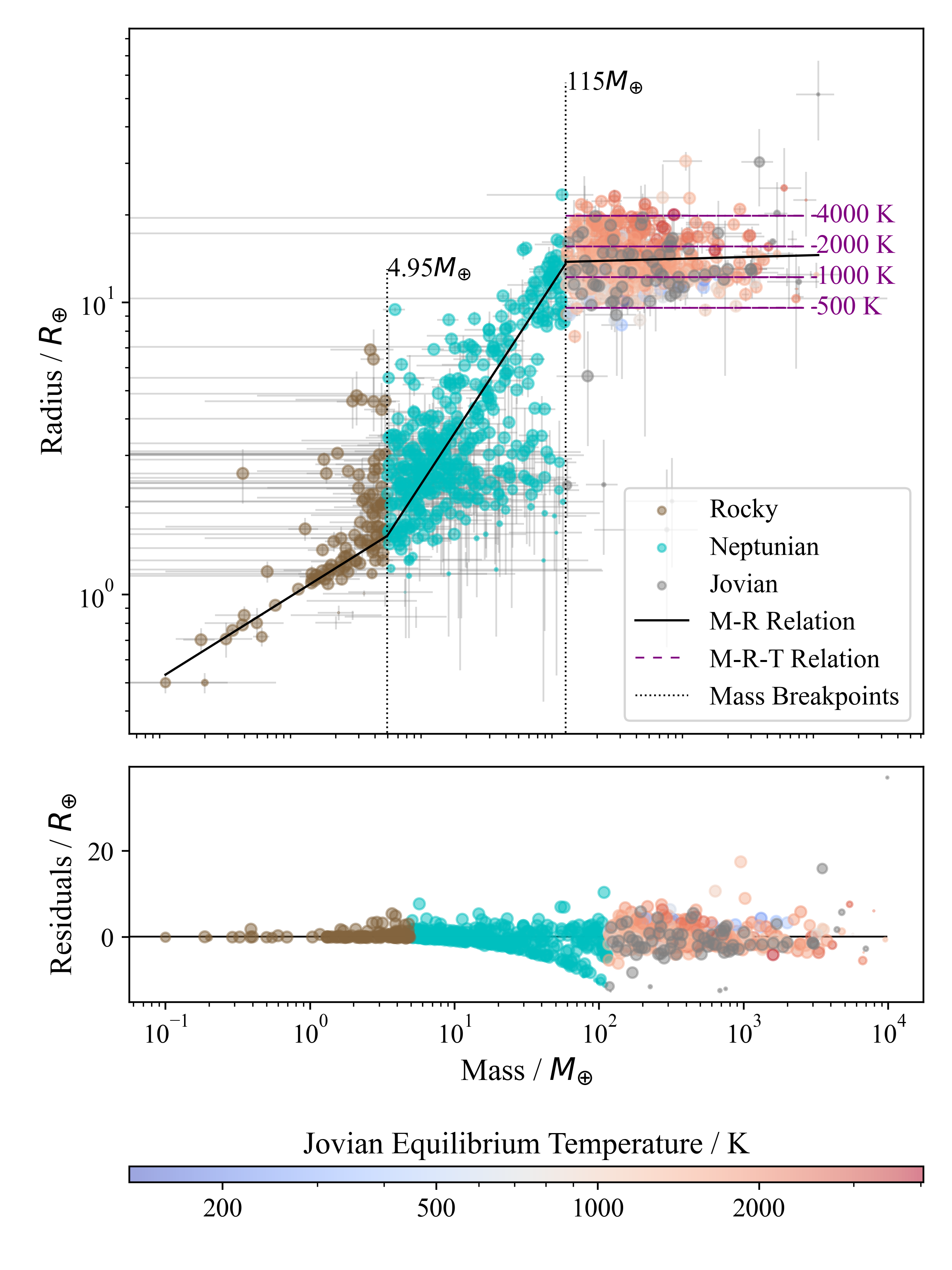}
\caption{A $M$-$R$ diagram in which planets have been classified into rocky, neptunian and jovian regimes according to the mass break-points $4.95 M_{\oplus}$ and $115 M_{\oplus}$. The black line indicates the three-piece $M$-$R$ relation found in section~\ref{piecewise}, and the contours of the $M$-$R$-$T$ relation in the jovian regime are plotted in purple. The size of the markers is proportional to the weighting factor. Rocky planets are coloured brown, neptunian planets are coloured cyan, and jovian planets follow a colour map according to their equilibrium temperature. jovian planets without an equilibrium temperature measurement are plotted in grey. The residuals are plotted in the lower panel, as the measured radius subtracted by the radius in the model. The average absolute difference between the model and measurements is 5.88.}
\label{hybrid_mr} 
\end{figure}

%\subsection{Continuous $M$-$R$-$T$ Model} \label{cont_mrt_model}

%Our results so far have provided mounting evidence that a mass break-point is not an ideal method for exoplanet classification, however the density classification scheme shown in Figure~\ref{broken_mr} leads to mom-unique radii at fixed mass within the rocky and neptunian regimes. As one of the primary purposes of a $M$-$R$ relation is prediction, 

Combining the continuous three-piece model from section~\ref{piecewise} with the $M$-$R$-$T$ model for jovian planets from Equation~\ref{eq:mrt} provides the semi-continuous $M$-$R$-$T$ model shown in Figure~\ref{hybrid_mr}.
The systematic downturn in the residuals of the neptunian regime due to misclassified rocky planets is once again evident as before in Figure~\ref{3_piece}. However, the scatter in jovian radii is reduced, with ${\cal M} = 5.88$,  illustrating the importance of modeling equilibrium temperature.

\subsection{Discontinuous mass-radius-temperature relations} \label{disc_mrt_model}

\begin{figure}
\includegraphics[width=0.45\textwidth]{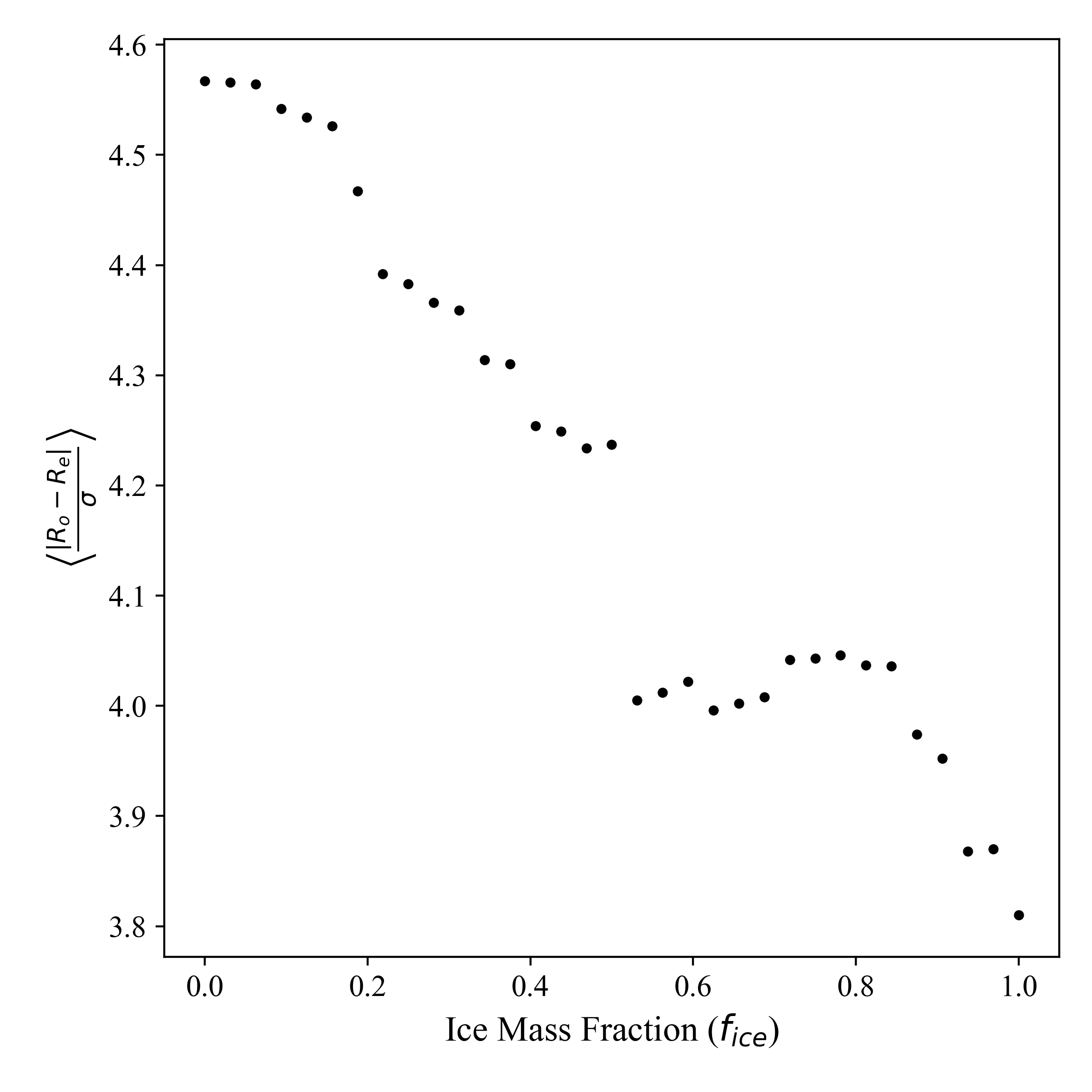}
\caption{A plot of the error-scaled average absolute difference between radius measurements and predictions [metric ${\cal M}$ in Equation~(\ref{r_metric})], against the assumed ice mass fraction $f_{\rm ice}$ used in Equation~\ref{imf_eos} to separate rocky from neptunian planets. A mass break-point of $M_{\rm br} = 115 M_{\oplus}$ is adopted to separate neptunian and  jovian planets.}
\label{imf} 
\end{figure}

The coherent nature of the residual excess seen in the neptunian regime in Figures~\ref{3_piece} and \ref{hybrid_mr} provides clear support for discontinuity in the transition from the rocky to neptunian regime. Whilst different planetary regimes can be segregated by specific mass or radius break points, we need a way to define distinct regions of the $M$-$R$ plane in order to consider discontinuous models \citep[c.f.][]{Otegi2020}.  To separate rocky from neptunian planets, we consider the ice/rock equation of state of \cite{Fortney2007}. The radius is given by
\begin{eqnarray}
\label{imf_eos}
R & = & (0.0592f_{\rm ice}+0.0975)(\log M)^2 \nonumber \\
  & &   +(0.2337f_{\rm ice}+0.4938)\log M \nonumber \\
  & &  +(0.3102 f_{\rm ice}+0.7932),
\end{eqnarray}
where $f_{\rm ice}$ is the ice mass fraction (1 for pure ice and 0 for pure rock), and mass and radius are in Earth units. We choose first to classify rocky planets as those which, for some fixed value of $f_{\rm ice}$, have a radius less than that calculated by substituting their mass measurement into Equation~\ref{imf_eos}. For the remaining planets, neptunian planets are defined as having a mass less than a mass break-point $M_{\rm br} = 115~M_{\oplus}$, while jovian planets have a mass larger than $M_{\rm br}$. After classification, discontinuous power law $M$-$R$ relations are fitted to the rocky and neptunian regimes, and a $M$-$R$-$T$ relation is fitted to the jovian regime as outlined in section~\ref{temp_inclusion}. We investigate the impact of our choice of $f_{\rm ice}$ using metric ${\cal M}$ defined by Equation~\ref{r_metric}. The dependence of ${\cal M}$ on $f_{\rm ice}$ is shown in Figure~\ref{imf}.

\begin{figure}
\includegraphics[width=0.45\textwidth]{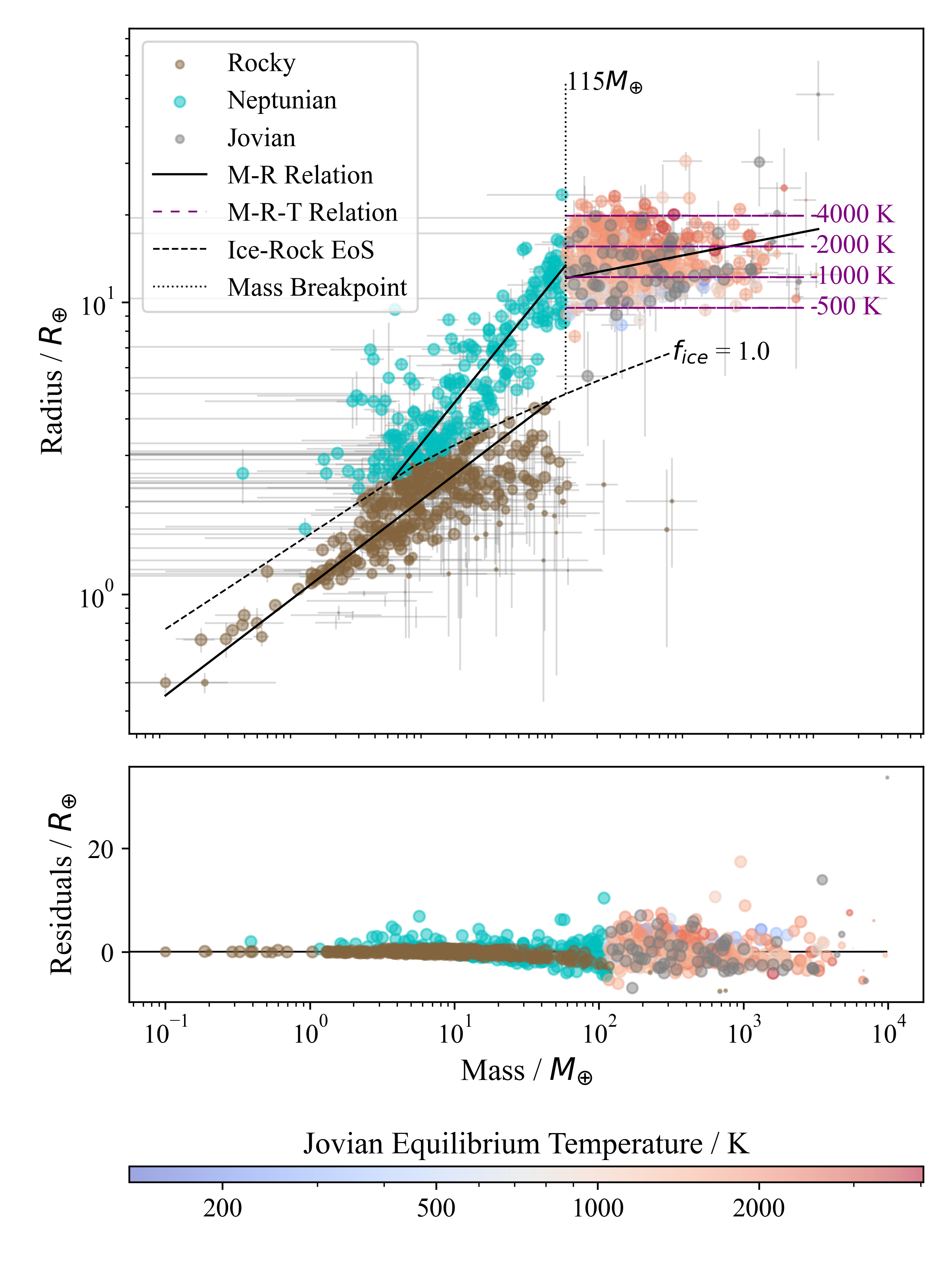}
\caption{A $M$-$R$ diagram in which planets have been classified into rocky, neptunian and jovian regimes according to the equation of state of a pure ice planet and a mass break-point of $115 M_{\oplus}$. The black lines indicate the power law $M$-$R$ relations fitted to the data, and fixed example contours of the $M$-$R$-$T$ relation in the jovian regime are plotted in purple. The size of the markers is proportional to the combined weighting factor. Rocky planets are coloured brown, neptunian planets are coloured cyan, and jovian planets follow a colour map according to their equilibrium temperature. jovian planets without an equilibrium temperature measurement are plotted in grey. The error-normalised residuals are plotted in the lower panel, with an average absolute difference of 3.81.}
\label{broken_mr} 
\end{figure}

In Figure~\ref{imf} we see a general downwards trend in ${\cal M}$, suggesting that a larger ice mass fraction is favourable for separating the super-Earth and mini-neptune regimes. Indeed, current data supports convergence to the approach of \cite{Otegi2020}, in which the composition line of water was used to separate planets. The observational discontinuity of the two regimes is consistent with a physical interpretation of a relatively sharp transition from a rock/ice to an ice giant regime. For our discontinuous $M$-$R$-$T$ relation we therefore adopt $f_{\rm ice} = 1$ to distinguish rocky planets from neptunes, and $M_{\rm br} = 115~M_{\oplus}$ to segregate neptunian and jovian planets.
The resulting $M$-$R$-$T$ relation is presented in Figure~\ref{broken_mr}.

From this model we find that when compared to Equation~\ref{power_law}, the rocky regime yields $k = 0.99 \pm 0.02$ and $\beta = 0.34 \pm 0.01$; the neptunian regime gives $k = 0.97 \pm 0.07$ and $\beta = 0.55 \pm 0.02$; and the jovian regime in the absence of equilibrium temperature data gives $k = 8.01 \pm 0.48$ and $\beta = 0.087 \pm 0.001$. Comparing to Equation~\ref{product_power}, the radii in the jovian regime are best described by $C = 1.10 \pm 0.15$ with temperature index $\beta_1 = 0.35 \pm 0.02$ and mass index $\beta_2 = 0.00\pm 0.01$. It is interesting that even the weak dependence of jovian radius on mass seen in Figure~\ref{3_piece} can apparently be explained away as pure temperature dependence.

\begin{figure*}
\includegraphics[width=\textwidth]{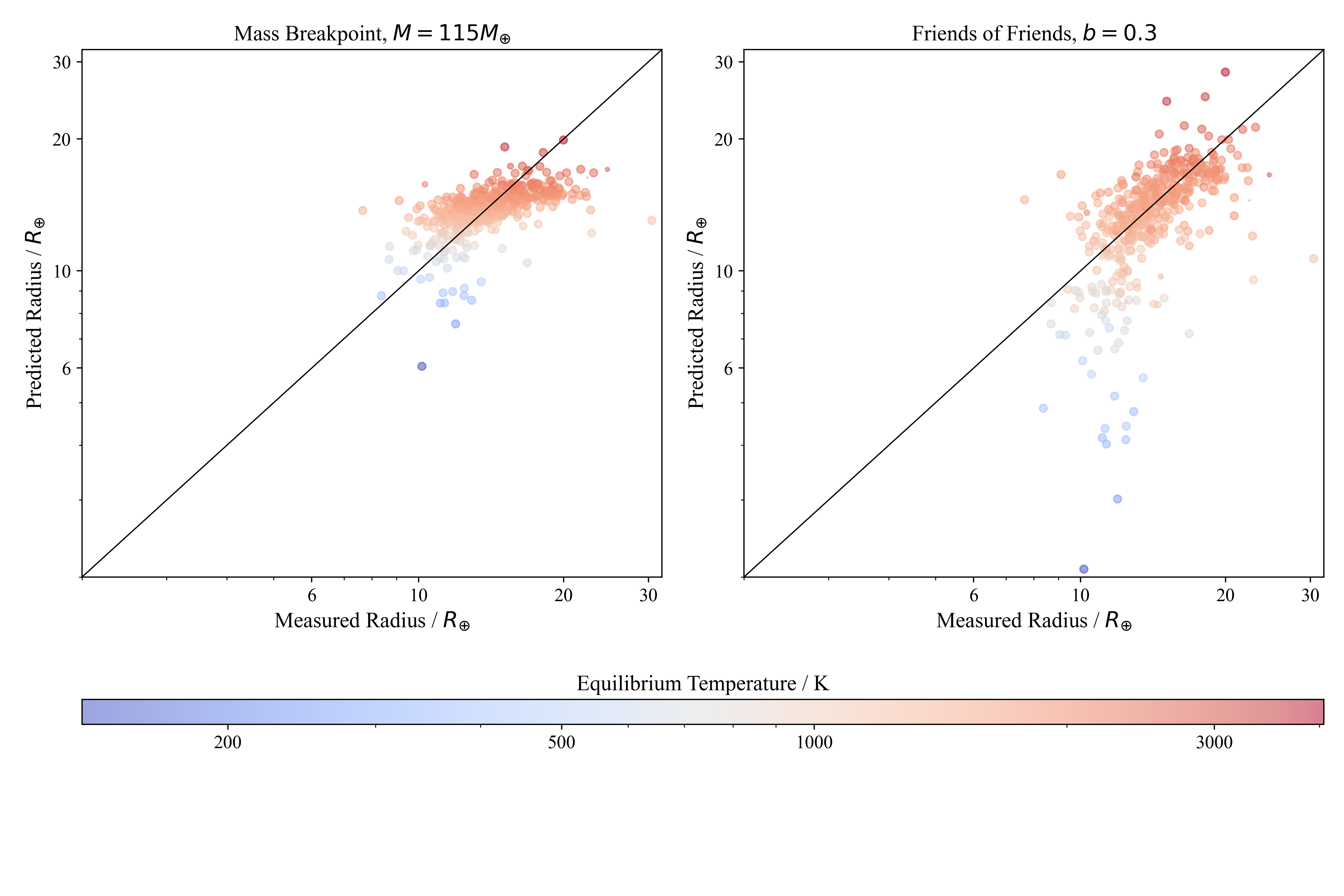}
\vspace{-1.5cm}
\caption{A plot of predicted radius against measured radius for jovian planets. The colour map represents the equilibrium temperature and the line for which the predicted radius is equal to the measured radius is plotted in black. On the left, the parameters for the model in Equation~\ref{product_power} are as plotted in Figure~\ref{broken_mr}. The planets shown have been selected using the mass break point $M_{\rm br} = 115~M_{\oplus}$. The distribution of hotter planets is clearly skewed with respect to the solid line demarcating perfect agreement between measured and predicted radii. On the right, we illustrate how one could correct for this by refitting only to the subset of planets selected using the friends-of-friends algorithm of \cite{Huchra1982}, with a clustering parameter, $b=0.3$. However, in this case we recover worsened predictions for the radii of cooler jovian worlds. }
\label{jovian_rr} 
\end{figure*}

In the residuals in Figure~\ref{broken_mr}, the radii of rocky planets appear to be well modelled, supported by the small uncertainty in the index of the power law in this regime. There remains some scatter in the neptunian regime, although the systematic down-turn from Figure~\ref{3_piece} is no longer present. Some scatter also remains in the jovian regime, and there is a slight downward trend in the residuals, which is made more apparent in Figure~\ref{jovian_rr}.

From the left panel in Figure~\ref{jovian_rr}, it appears that the model is skewed away from the visible trend by some cooler planets that have radii larger than expected. This is likely an effect of implementing a break-point, as the transition between neptunian and jovian planets may itself be temperature sensitive. 
%{\color{red} {\bf Comment:} might this be fixable by allowing for the $M$-$R$-$T$ relation of Eq~\ref{eq:mrt} to extend down into the neptunian regime? Can we do a continuous two-piece $M$-$R$-$T$ planar fit, with either a $T$-independent $M_{\rm br}$ parameter or one that itself varies smoothly as a power of $T$, to model all planets lying above the $f_{\rm ice} = 1$ line? This might just solve the issue highlighted in Figure~\ref{jovian_rr}, as well as reduce the scatter for neptunian worlds.}
%% Removing this statement is it is not true - there are hot bloated neptunes.
%Neptunian planets with large volatile envelopes are likely to reside at larger host separations, otherwise insolation from the host star would evaporate their envelopes to leave a rocky planet. Hence, the radii of these planets do not respond to temperature in the same way as jovian radii therefore skewing the fit. 
The right panel of Figure~\ref{jovian_rr} uses the friends-of-friends algorithm of \cite{Huchra1982} with a clustering parameter, $b=0.3$, to isolate the main group of hotter jovian planets in $M$-$R$-$T$ space. We then use this data to refit parameters in the $M$-$R$-$T$ model. This new model is used to predict the radii of the same planets in the left panel for direct comparison. As this approach essentially removes outlying planets, the predicted radii correlate better with the measured radii for the bulk of the jovian planets. However, it does worsen the predictions for cooler planets, though some of these may well be misclassified neptunian planets. A fixed mass break-point between neptunes and jovian planets is well-motivated by the general trend, but may not be optimal.

\section{Comparison of relations} \label{comparison}

\begin{figure}
\includegraphics[width=0.45\textwidth]{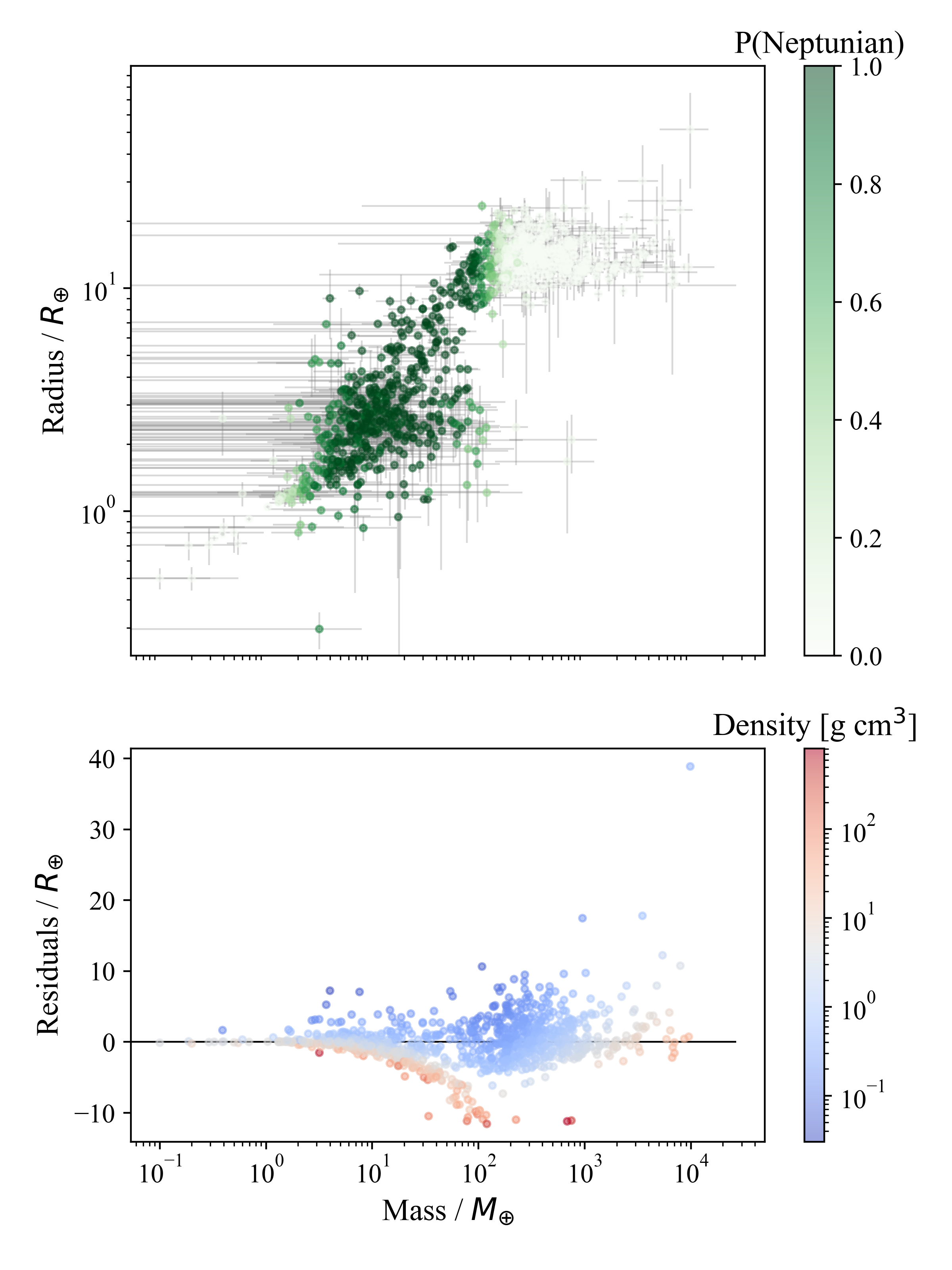}
\caption{A plot of radius against mass in the upper panel, where the colour corresponds to the probability \texttt{Forecaster} has assigned to each planet of being neptunian. The residuals plotted in the lower panel were calculated as the measured radii subtracted by the predicted radii generated from a single run of \texttt{Forecaster}, with the colour representing the bulk density of each planet.}
\label{forecaster_mr} 
\end{figure}

\texttt{Forecaster} is a widely-used, publicly-available program for predicting exoplanet masses and radii on a probabilistic basis, developed by \cite{ChenKipping2017}. For a given mass or radius with optional uncertainties, \texttt{Forecaster} returns a prediction of the other measurement based on how likely the planet is to fall into each regime and the uncertainty regions surrounding the model. As this model is probabilistic, a different prediction value will be returned each time the program is run. The terran, neptunian, jovian and stellar regimes are split by mass break-points, and a continuous broken power law relates mass and radius, where each segment of the power law has a different uncertainty region. Figure~\ref{forecaster_mr} shows the $M$-$R$ diagram coloured according the the probability \texttt{Forecaster} has assigned to each planet of being neptunian. The residuals for one run of \texttt{Forecaster} are also shown and coloured according to their bulk density.

Figure~\ref{forecaster_mr} demonstrates that \texttt{Forecaster} suffers from the same systematic downturn in the residuals of the neptunian regime as our three-piece model in Figure~\ref{3_piece}. From the residuals panel, it is also apparent that these planets with over-predicted radii are among the densest planets, supporting the presumption that these are misclassified rocky planets. This is once again a feature of using a mass break-point to distinguish between rocky and neptunian planets. Additionally, a large scatter in radii is evident in the residuals for large masses due to an absence of temperature dependence in the model.

The slope of a log-log $M$-$R$ plot for our three-piece model is consistent with those found both by \cite{ChenKipping2017} and by \cite{Otegi2020} in the rocky regime; all around 0.28. However, our discontinuous model gives a steeper slope of $0.34 \pm 0.01$ in this regime, consistent with $R \propto M^{\frac{1}{3}}$ as expected for a solid body with constant density across all rocky planets. This contrasts with \cite{Schulze2021} in that it argues for a rocky planet bulk density that is, on average, insensitive to the composition of its host. In the neptunian regime, the three-piece model gives a slope larger than that used by \texttt{Forecaster}, potentially due to the smaller break-point found by \cite{ChenKipping2017} to separate between rocky and neptunian planets. Our discontinuous model, on the other hand, gives a slope smaller than that found by both \cite{Otegi2020}, and \cite{ChenKipping2017}. The difference from \texttt{Forecaster} can be attributed to the use of a density-based classification scheme rather than a mass break-point, however this was also the approach taken by \cite{Otegi2020}. This difference in slope may arise from the subtly different cutoff for higher mass planets of $120 M_{\oplus}$ compared to $115 M_{\oplus}$ used in our model, as well as from differences in the data used and in fitting approaches. 

The slopes of the jovian $M$-$R$ relation found by \cite{ChenKipping2017} and our three-piece and discontinuous models are not consistent with one another. We expect that this is due to the large scatter in radii and weak mass-dependence, hence subtle changes to the exact data included or the location of the break-point can have a significant impact on the results of the fit. Furthermore, \cite{ChenKipping2017} included brown dwarfs in their sample, which anchor the fit behaviour within the jovian regime. Nonetheless, all of the $M$-$R$ slopes are very shallow and therefore near degenerate. In our discontinuous $M$-$R$-$T$ model, we find that $M$-$T$ dependence has a slope of 0.35 $\pm$ 0.02, but that $M$ is essentially uncorrelated to $R$, with a slope of $0.00\pm 0.01$.

The use of a density-based classification scheme has the interesting consequence that planets we classify to be rocky have masses up to $79~M_{\oplus}$, comparable to relatively high mass neptunian bodies. Whilst \cite{ChenKipping2017} found that data at the time indicated that rocky super-Earths were not nearly as large as expected, current data indicates rocky planets can be as large as $4.3~R_{\oplus}$.
\begin{figure*}
\includegraphics[width=\textwidth]{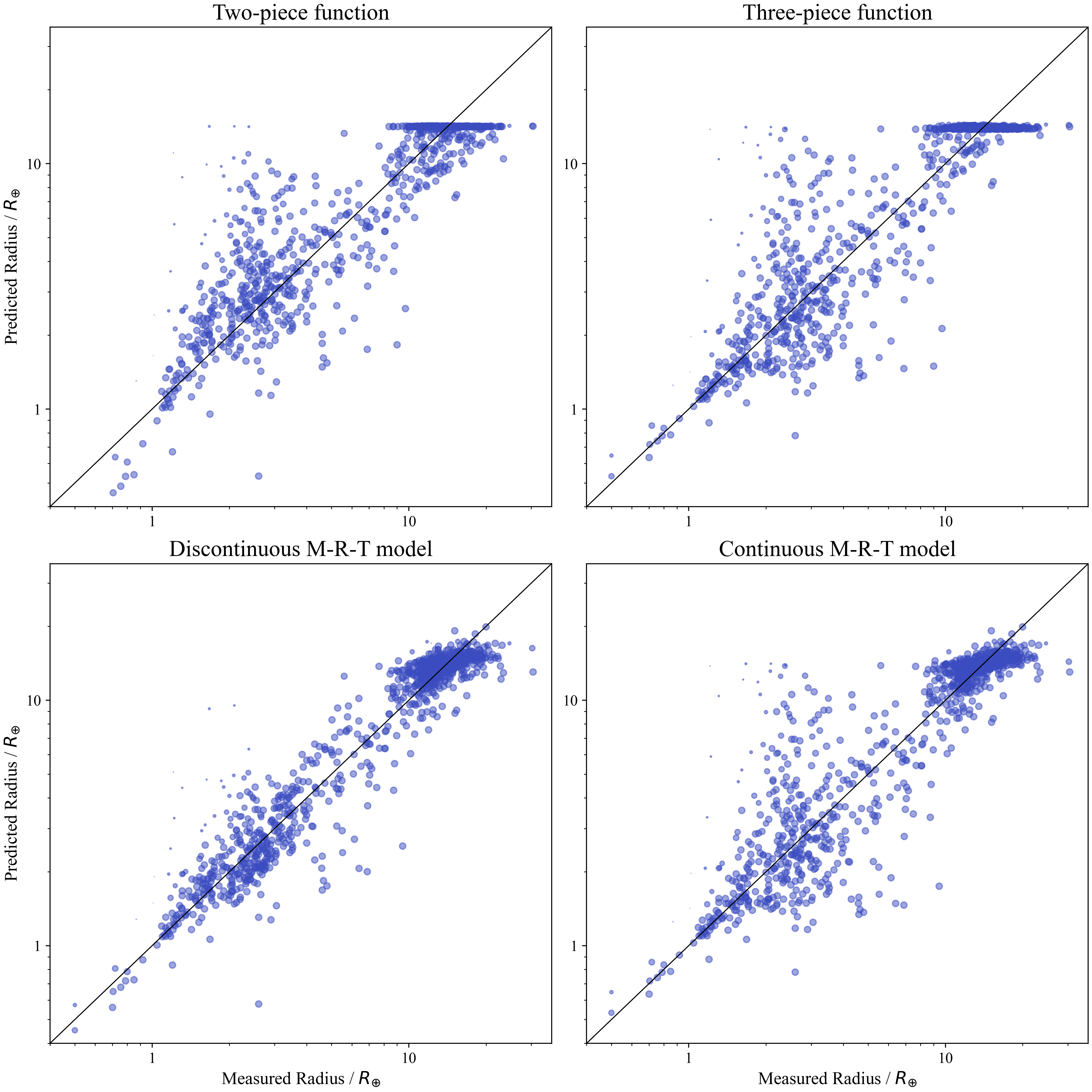}
\caption{Plots of the radii predicted by a two-piece function, three piece function, discontinuous $M$-$R$-$T$ model and continuous $M$-$R$-$T$ model against measured radius. The size of the markers is proportional to the combined weighting factors used to down-weight points in the ODR fit which did not satisfy the plausibility criteria. The line for which the predicted radius is equal to the measured radius is plotted in black.}
\label{rr_plots} 
\end{figure*}

A plot of predicted radii against measured radii for the four models we have developed are displayed in Figure~\ref{rr_plots}.
The 2-piece relation in Figure~\ref{rr_plots} shows significant deviation from the unity line for the smallest planets, while for the other three models agreement is much stronger, confirming that the data provides strong support for three distinct planetary regimes. Both $M$-$R$ models in Figure~\ref{rr_plots} show a systematic degeneracy between measured and predicted radii for the largest planets. There is much better correlation of predictions with measurements for the $M$-$R$-$T$ models. All continuous models show large scatter around the unity line for intermediate mass planets due to the superposition of the super-Earth and mini-neptune regimes. This scatter is greatly reduced by the discontinuous $M$-$R$-$T$ model.
\begin{figure}
\includegraphics[width=0.45\textwidth]{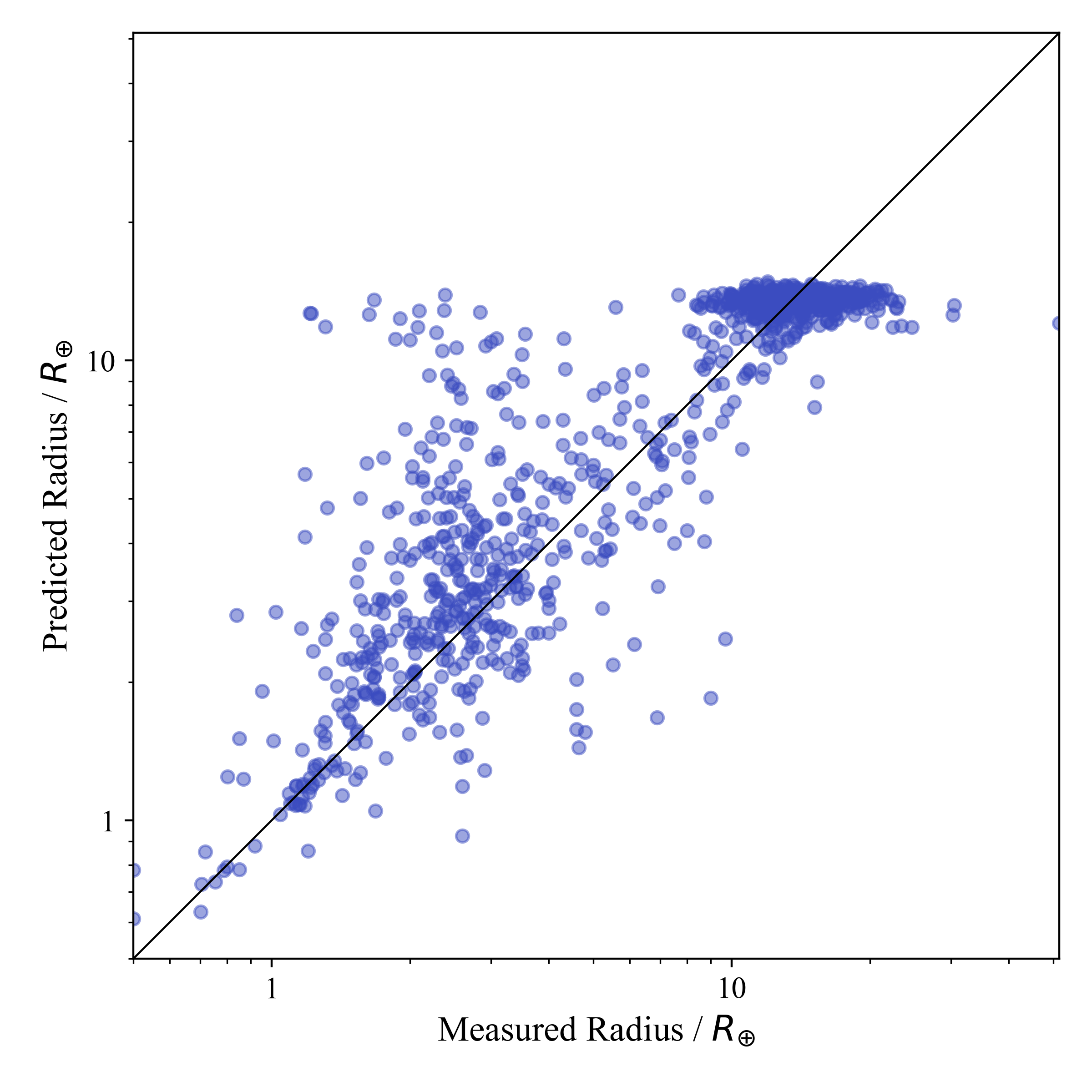}
\caption{A plot of the radius predicted by one run of \texttt{Forecaster} against measured radius. The line for which the predicted radius is equal to the measured radius is plotted in black.}
\label{forecaster_rr} 
\end{figure}

The large scatter for intermediate mass planets is similarly seen in Figure~\ref{forecaster_rr}, which shows the results of predicted versus measured radii for a single run of \texttt{Forecaster}. Additionally, whilst a probabilistic model goes some way towards reproducing the radius scatter of large planets, it does not perform as well as a model that includes temperature to account for bloating.

\begin{figure*}
\includegraphics[width=\textwidth]{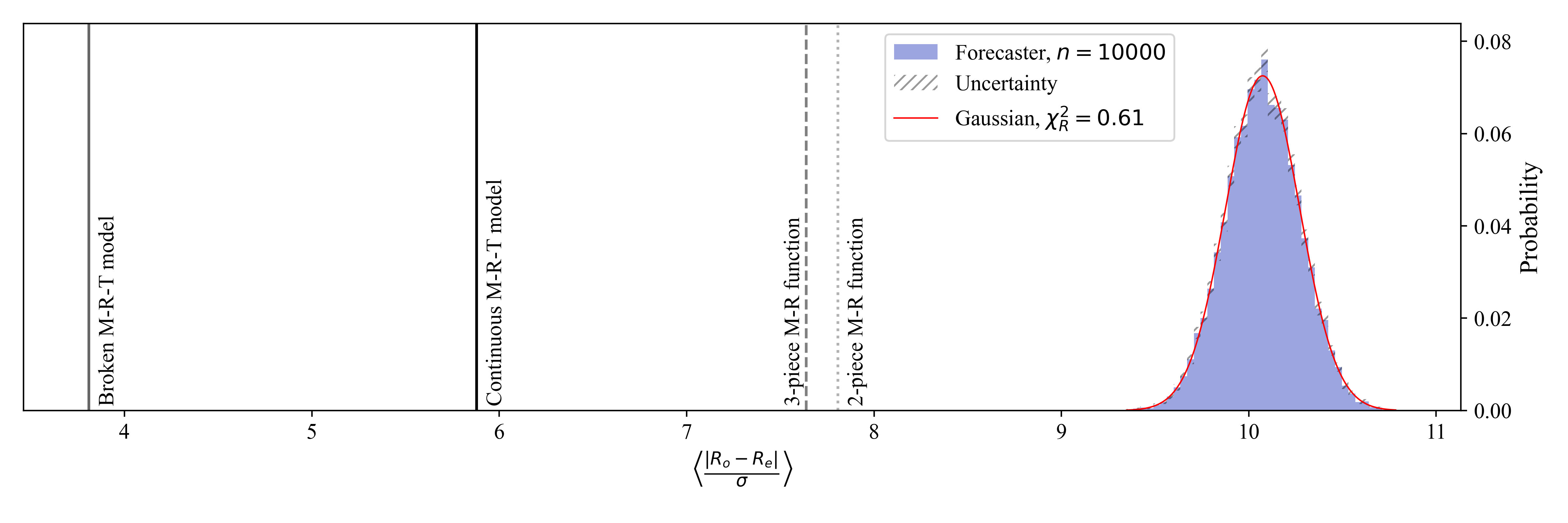}
\caption{Metric ${\cal M}$ calculated for our two- and three-piece $M$-$R$ models as well as our continuous and discontinuous $M$-$R$-$T$ models applied to the current data set. The distribution of ${\cal M}$ for 10,000 runs of \texttt{Forecaster} has been fitted to a Gaussian distribution, with $\langle {\cal M} \rangle = 10.1$ and $\sigma({\cal M}) = 0.2$.}
\label{comparison_new} 
\end{figure*}

We can use Equation~\ref{r_metric} to quantitatively assess the success of each model via ${\cal M}$. Since \texttt{Forecaster} is a probabilistic model, ${\cal M}$ will differ for each run of the program. We therefore opt to calculate it for 10,000 runs and to fit a Gaussian to the resulting distributions. The results are displayed in Figure~\ref{comparison_new}.

Figure~\ref{comparison_new} clearly demonstrates that when considering the current data available, there is no case in which \texttt{Forecaster}'s average performance is better than that of our deterministic models. $M$-$R$-$T$ models outperform $M$-$R$ models and a discontinuous model is favoured over a continuous one.

One reason why \texttt{Forecaster} may not perform well compared to our models may simply be because it is conditioned on much older data. At the time of writing, the public version of \texttt{Forecaster}\footnote{\url{https://github.com/chenjj2/forecaster}} is calibrated upon data available at the time of its initial development, which is prior to 2017.  To fairly compare models, we consider how our models predict data available to \cite{ChenKipping2017}, though with some caveats.

As we are interested in an exoplanet mass-radius relation, we neglect any measurements for objects that are classified as stars or brown dwarfs. Furthermore, like \cite{ChenKipping2017} we include solar system bodies in our sample, but we use current measurements and uncertainties taken from the JPL Planetary Satellite Physical Parameters\footnote{\url{https://ssd.jpl.nasa.gov/sats/phys_par/}} database, \cite{Williams2014} and the JPL Planetary Physical Parameters\footnote{\url{https://ssd.jpl.nasa.gov/planets/phys_par.html}} database. The full table of solar system parameters is presented in Table~\ref{ss_parameters} of the Appendix, along with our equilibrium temperature calculation for Jupiter. We use these updated values both within \texttt{Forecaster} and for our own relations for consistency. The uncertainties for these bodies are small, hence we do not expect their revision to significantly alter the performance of \texttt{Forecaster}.

\begin{figure}
\includegraphics[width=0.48\textwidth]{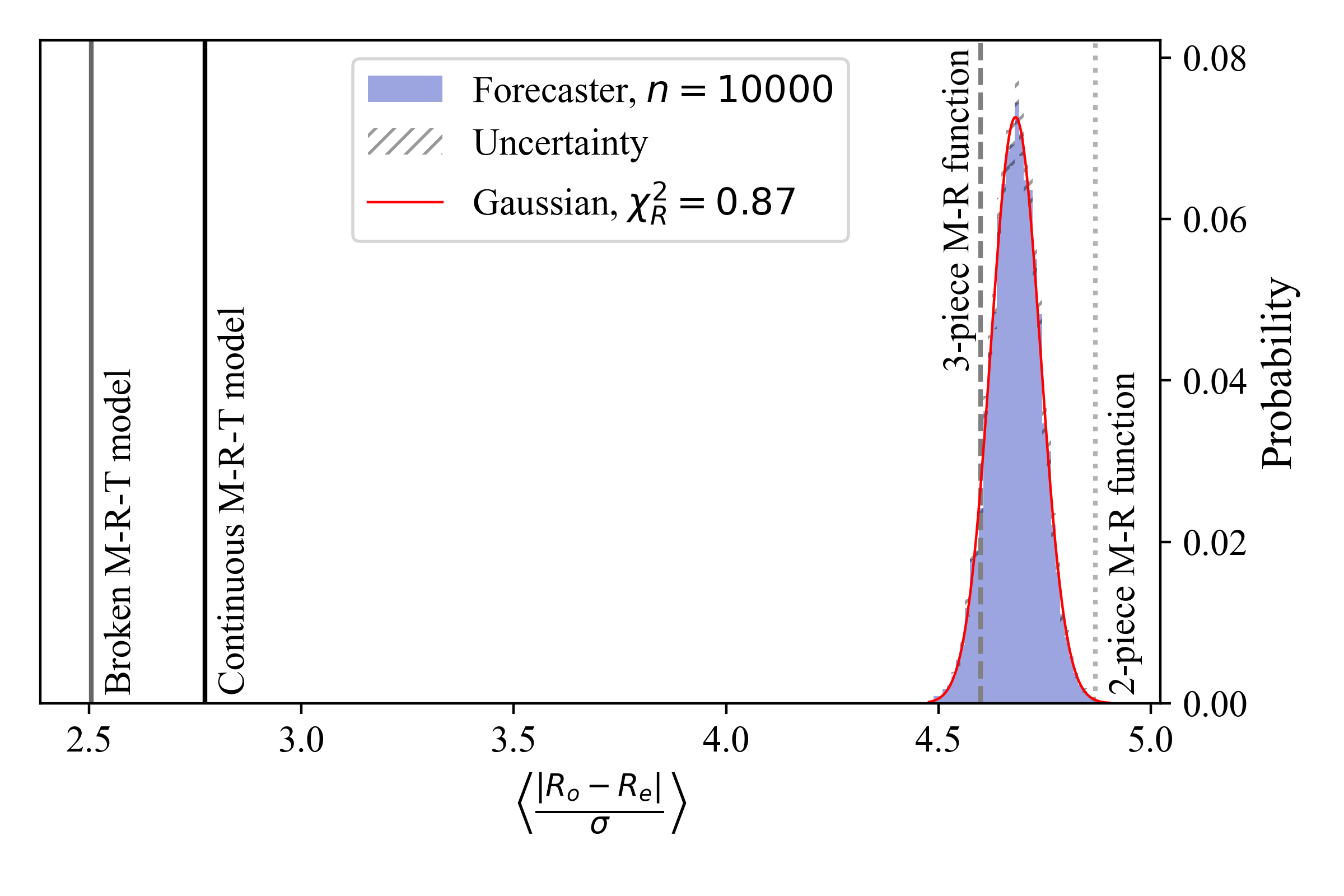}
\caption{Metric ${\cal M}$ for our two- and three-piece $M$-$R$ models as well as our continuous and discontinuous $M$-$R$-$T$ models applied to the old data set without the inclusion of solar system bodies. The distribution of ${\cal M}$ for 10,000 runs of \texttt{Forecaster} has been fitted to a Gaussian distribution with $\langle {\cal M} \rangle = 4.68$ and $\sigma({\cal M}) =  0.06$.}
\label{comparison_old} 
\end{figure}

\begin{figure}
\includegraphics[width=0.45\textwidth]{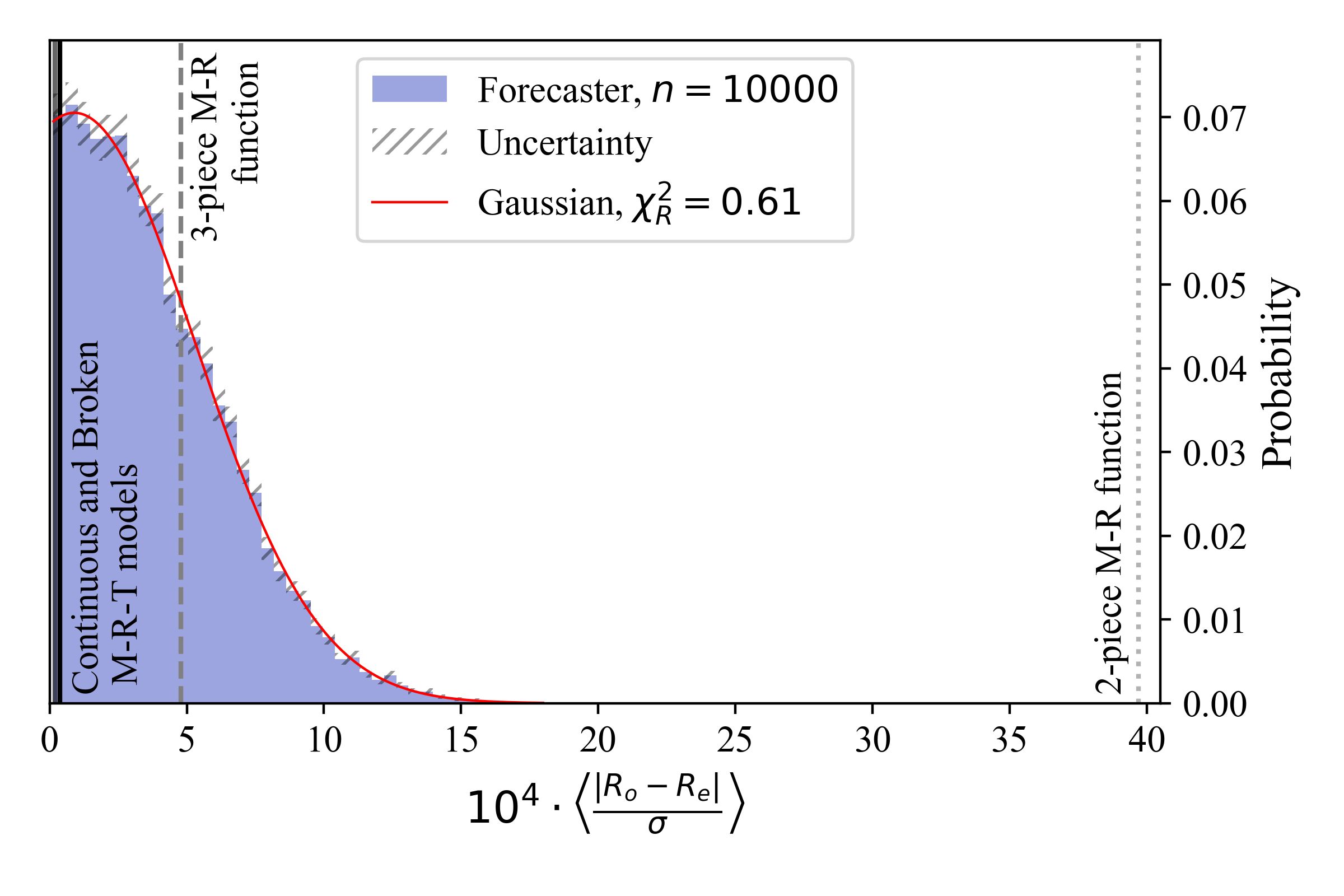}
\caption{Metric ${\cal M}$ for our two- and three-piece $M$-$R$ models as well as our continuous and discontinuous $M$-$R$-$T$ models applied to solar system bodies. The ${\cal M}$ axis is four orders of magnitude larger than in previous similar plots and so the metrics calculated for the $M$-$R$-$T$ models cannot be resolved in this figure. The distribution of ${\cal M}$ for 10,000 runs of \texttt{Forecaster} has been fitted to a Gaussian with ${\cal M} = 9000$ and $\sigma({\cal M}) = 44000$. None of the models is able to predict the radii of solar systems bodies to within their very small current uncertainties, though the $M$-$R$-$T$ models perform much better, in relative terms, than the other models.}
\label{comparison_ss} 
\end{figure}

We compare ${\cal M}$ for the dataset subset that existed at the time of \cite{ChenKipping2017} in Figure~\ref{comparison_old}, with solar system objects excluded. This leaves only one exoplanet in the old data set that \texttt{Forecaster} classifies as rocky, and this lies within the uncertainty region of the $2 M_{\oplus}$ break-point. As a result, \texttt{Forecaster} has to rely on the inclusion of solar system values in this regime. In Figure~\ref{comparison_ss} we show the result of ${\cal M}$ for the model predictions of the size of solar system bodies.

In Figure~\ref{comparison_old} we see that while our $M$-$R$-$T$ models still significantly outperform \texttt{Forecaster}, our $M$-$R$ models are now comparable. Using the mean and width of the Gaussian fitted to the \texttt{Forecaster} ${\cal M}$ distribution, we find that \texttt{Forecaster} can outperform the two-piece $M$-$R$ model 99.9\% of the time, but can out-perform our three-piece $M$-$R$ model only 8\% of the time. 

We see in Figure~\ref{comparison_ss} that our two-piece model performs comparatively poorly for predicting the size of solar system bodies. Our other models are able to extrapolate more reliably down to this low-mass regime though, unsurprisingly, nowhere near the precision of current measurement uncertainty for these bodies. We find that \texttt{Forecaster} outperforms a three-piece $M$-$R$ relation 67\% of the time, a continuous $M$-$R$-$T$ model 6\% of the time and a discontinuous $M$-$R$-$T$ model 3\% of the time. We conclude that a $M$-$R$-$T$ model calibrated on current data nearly always outperforms \texttt{Forecaster} calibrated on data prior to 2017. We expect that the performance of \texttt{Forecaster} could be significantly improved by updating the hyper-parameters of the \cite{ChenKipping2017} model to include current measurements, though the means to do so have not been made publicly available. Nonetheless, the weaknesses of \texttt{Forecaster} are also those inherent to a continuous $M$-$R$ model with mass break-points, and as such are unlikely to be fully mitigated by updating the input dataset.

One aspect that has not been accounted for in any of the models discussed in this report is the effect of detection and selection bias on the mass-radius dataset. With the exception of \cite{Burt2018}, prioritisation schemes for the follow-up of exoplanet detections are not generally made available and there are very few investigations into how these schemes may introduce bias into the population of planets for which we have both a mass and a radius measurement. This therefore makes it very difficult to de-bias $M$-$R$ relations calibrated on observations.

As for detection bias, the densities of planets calculated using transit timing variations (TTV) tend be smaller than those of planets with mass measurements obtained from radial velocities. \cite{Leleu2022} has found evidence to suggest that some of this discrepancy is due to differing detection biases, and they correct these measurements accordingly. A similar attempt to account for biased TTV planet measurements is presented in \cite{Hutter2020}. The number of TTV planets in our sample is however vastly out-numbered by the number of planets meaaured using radial velocities, so we expect any TTV biasing effect to be small.

\section{Conclusions} \label{conclusions}

We have compiled a catalogue of 1053 confirmed exoplanets with which to calibrate power-law exoplanet mass-radius ($M$-$R$) and mass-radius-temperature ($M$-$R$-$T$) relationships. We have strived to let the data itself inform us as to the piece-wise structure of these relationships, including whether continuous or discontinuous power-law forms are preferred.

Using orthogonal distance regression fits that account for errors in both mass and radius, we find that current data is best explained by three distinct planetary regimes that, under a continuous $M$-$R$ relation, transition from a rocky to an ice giant (neptunian) regime at $4.95\pm 0.81~M_{\oplus}$ and from ice giant to gas giant (jovian) at $115~M_{\oplus}$. 

We find that the modeling of the jovian regime is improved through inclusion of the effect of bloating via extension to an $M$-$R$-$T$ relation. In fact, when doing this, we find that $M \propto R^{0.00\pm 0.01} T^{0.35\pm 0.02}$, so that jovian mass planets can be well modeled with no radius dependence at all.

Our analysis also finds strong support from the data for a discontinuous $M$-$R$-$T$ relation between rocky and neptunian planets, as has been previously argued by \cite{Otegi2020}. Modeling the boundary with analytic ice-rock equations of state from \cite{Fortney2007} we find that the data prefers a boundary corresponding to a pure-ice world, giving support for the physical interpretation of the discontinuity as separating rocky from ice-giant (neptunian) planet populations. Interestingly, we find that the resulting upper mass of planets categorized within the rocky planet regime can extend almost up to the upper mass limit of the neptunian population.

Given the significant increase in the amount of exoplanet data since the publication of the widely-used \texttt{Forecaster} code \citep{ChenKipping2017} we find most of our models outperform \texttt{Forecaster} in the accuracy of radius predictions. While this can to an extent be attributed to the hyper-parameters of the \texttt{Forecaster} model being conditioned on an older and smaller dataset, the models which perform the best against it are those that allow for discontinuities arising from variations in temperature and equation of state that are not included in the underlying $M$-$R$ model used in \texttt{Forecaster}.

Looking ahead, the current exoplanet dataset will see a massive expansion over the coming decade, thanks largely to astrometric detection by ESA Gaia \citep{Perryman2014} and by the transit and microlensing samples of the NASA Nancy Grace Roman Space Telescope \citep{2019ApJS..241....3P,2023arXiv230516204W}. Roman alone is expected to expand the exoplanet catalogue from the current size of under 6,000 planets to at least 60,000 and possibly up to 200,000 hot transiting planets, as well around 1,400 cool microlensing planets. As these datasets will involve combinations of mass (astrometry and microlensing) and radius (transit) measurements, a coherent analysis of exoplanet demography will require an increasingly precise modelling of the exoplanet $M$-$R$-$T$ relation.

\section*{Acknowledgements}

This research has made use of the NASA Exoplanet Archive, which is operated by the California Institute of Technology, under contract with the National Aeronautics and Space Administration under the Exoplanet Exploration Program.

\bibliographystyle{mnras}
\bibliography{references}

\section*{Appendix} \label{ss_params}

The only solar system object with a mass greater than $115 M_{\oplus}$ and thus requiring an equilibrium temperature measurement is Jupiter. Assuming that both the Sun and Jupiter behave as blackbodies, it can be shown that the equilibrium temperature of Jupiter, $T_{\rm eq,J}$, is
\begin{eqnarray}
\label{T_eq}
T_{\rm eq,J} = T_{\odot} \left( \frac{R_{\odot}}{2a}\right)^{\frac{1}{2}} (1-A)^{\frac{1}{4}},
\end{eqnarray}
where $T_{\odot} = 5772$~K is the solar effective temperature, $R_{\odot}$ is the solar radius, $a = 5.2$~au is the separation between Jupiter and the Sun, and $A = 0.5$ is the Bond albedo of Jupiter. The substitution of these values into Eqn~(\ref{T_eq}) and propagating their uncertainties yields an equilibrium temperature for Jupiter of $102.3 \pm 0.6$ K.

The masses and radii of solar system objects are presented in Table~\ref{ss_parameters}.
\begin{table}
%\begin{ruledtabular}
\caption{The solar system objects considered in \texttt{Forecaster} with updated values for their masses and mean radii including uncertainties. Parameters for moons are taken from the JPL Planetary Satellite Physical Parameters database, excepting the mass of the Moon, which was taken from \cite{Williams2014}. All planet parameters are taken from the JPL Planetary Physical Parameters database.}
\begin{tabular}{lll}
\textbf{Object} & \textbf{Mass (kg)} & \textbf{Mean Radius (km)}\\
\hline
\textsc{Moons} & & \\
Moon & (7.3463 $\pm$ 0.0088) $\cdot 10^{22}$ & 1737.4 $\pm$ 0.1\\
Io & (8.931938 $\pm$ 0.000018) $\cdot 10^{22}$ & 1821.49 $\pm$ 0.5\\
Europa & (4.799844 $\pm$ 0.000013) $\cdot 10^{22}$ & 1560.8 $\pm$ 0.3\\
Ganymede & (1.4819 $\pm$ 0.0001) $\cdot 10^{23}$ & 2631.2 $\pm$ 1.7\\
Callisto & (1.07594 $\pm$ 0.00014) $\cdot 10^{23}$ & 2410.3 $\pm$ 1.5\\
Rhea & (2.306520 $\pm$ 0.000035) $\cdot 10^{21}$ & 2410.3 $\pm$ 1.5\\
Titan & (1.3452 $\pm$ 0.0002) $\cdot 10^{23}$ & 2574.76 $\pm$ 0.02\\
Titania & (3.400 $\pm$ 0.061) $\cdot 10^{21}$ & 788.9 $\pm$ 1.8\\
Oberon & (3.076 $\pm$ 0.087) $\cdot 10^{21}$ & 761.4 $\pm$ 2.6\\
Triton & (2.1390 $\pm$ 0.0028) $\cdot 10^{22}$ & 1352.6 $\pm$ 2.4\\
\hline
\textsc{Dwarf Planets} & & \\
Eris & (1.660 $\pm$ 0.020) $\cdot 10^{22}$ & 1200 $\pm$ 50\\
Pluto & (1.3029	$\pm$ 0.0027) $\cdot 10^{22}$ & 1188.3 $\pm$ 1.6\\
\hline
\textsc{Planets} & & \\
Mercury & (3.30103 $\pm$ 0.00021) $\cdot 10^{23}$ & 2439.4 $\pm$ 0.1\\
Venus & (4.86731 $\pm$ 0.00023) $\cdot 10^{24}$ & 6051.8 $\pm$ 1.0\\
Earth & (5.97217 $\pm$ 0.00028) $\cdot 10^{24}$ & 6371.0080 $\pm$ 0.0001\\
Mars & (6.41691 $\pm$ 0.00030) $\cdot 10^{23}$ & 3389.5 $\pm$ 0.2\\
Jupiter & (1.898125 $\pm$ 0.000088) $\cdot 10^{27}$ & 69911	$\pm$ 6\\
Saturn & (5.68317 $\pm$ 0.00026) $\cdot 10^{26}$ & 58232 $\pm$ 6\\
Uranus & (8.68099 $\pm$ 0.0004) $\cdot 10^{25}$ & 25362 $\pm$ 7\\
neptune & (1.024092 $\pm$ 0.000048) $\cdot 10^{26}$ & 24622	$\pm$ 19\\
\end{tabular}
\label{ss_parameters}
%\end{ruledtabular}
\end{table}
\end{document}